\documentclass{article}   	

\usepackage{graphicx}						
\usepackage{amssymb}
\usepackage{amsmath}
\usepackage{mleftright}

\usepackage{physics}
\usepackage{autonum}
\usepackage{float}

\usepackage{cite}

\usepackage[T2A]{fontenc}
\usepackage[utf8]{inputenc}

\usepackage[titletoc,title]{appendix}

\begin{document}

\begin{titlepage}
 	  \begin{center}

    		   \huge{Does the entropy of systems with larger internal entanglement grow stronger?}

			\vspace{1cm}

		\Large{Daria Gaidukevich}
			\vspace{0.3cm}

		\large{Independent researcher}
		\vspace{0.3cm}
		
		\large{dasha.gaidukevich@gmail.com}
		\vspace{0.3cm}

 	  \end{center}

\abstract{
It is known that when a system interacts with its environment, the entanglement contained in the system is redistributed since parts of the system entangle with the environment. On the other hand, the entanglement of a system with its environment is closely related to the entropy of the system. However, does this imply that the entropy of systems with larger internal entanglement will grow stronger?
We study the issue using the simplest model as an example: а system of qubits interacts with the environment described by the quantum harmonic oscillator. The answer to the posed question is ambiguous. However, the study of the situation on average (using the simulation of a set of random states) reveals certain patterns and we can say that the answer is affirmative. At the same time, the choice of states satisfying certain conditions in some cases can change the dependence to the opposite.  Additionally, we show that the entanglement depth also makes a small contribution to entropy growth.
} 

\vspace{0.5cm}

 
 \end{titlepage}

\section{Introduction}

The connection between the second law and quantum correlations plays more than a significant role in modern physics. The relationship between entropy, entropy change or entropy production, and quantum correlations is discussed in numerous articles and can be considered from different points of view 
 \cite{Entropy production as correlation,Quantum Entanglement and Entropy,Quantum mechanical evolution,Interplay between entanglement and entropy,quantum correlations / violation of the second law,Entanglement and the foundations of statistical mechanics,Entropy production and correlations in a controlled non-Markovian setting,Maxwell's demon,Entropy production in open systems,entropic criterion for separability,Quantum correlation entropy,Entropy Production in Non-Markovian Collision Models,Maximal entanglement versus entropy,Dynamics of quantum entanglement, Production of entanglement entropy by decoherence, Deutsch, Q and class correlations,Entropy dynamics in the system of interacting qubits,Quantum measurements and equilibration: the emergence of objective reality via entropy maximisation,Correlations in quantum thermodynamics: Heat work and entropy production,Entanglement versus Stosszahlansatz: Disappearance of the thermodynamic arrow in a high-correlation environment,Entropy production and the role of correlations in quantum Brownian motion,Fluctuation theorem with information exchange: Role of correlations in stochastic thermodynamics}). For instance, one can inquire about the mutual dynamics of internal entanglement and entropy of a system \cite{Dynamics of quantum entanglement,Interplay between entanglement and entropy}, ensure that quantum correlations do ``not lead to violation of the second law of thermodynamics'' \cite{quantum correlations / violation of the second law}, think about the role of entanglement in equilibration process \cite{Entanglement and the foundations of statistical mechanics,Quantum mechanical evolution}, relate thermodynamic entropy to the entanglement between parts of a large system \cite{Deutsch} and use entropy production as a measure of system-reservoir correlations \cite{Entropy production as correlation}. From any perspective, the issue appears to be interesting and potentially useful both for specific areas of physics (like 
quantum information theory \cite{Entropy dynamics in the system of interacting qubits,entropic criterion for separability}, 
quantum thermodynamics \cite{Correlations in quantum thermodynamics: Heat work and entropy production,Quantum mechanical evolution},
 open systems \cite{Entropy production in open systems, Entropy production and the role of correlations in quantum Brownian motion},
  non-Markovian systems \cite{Entropy production and correlations in a controlled non-Markovian setting, Entropy Production in Non-Markovian Collision Models}) and for the general understanding of the thermodynamics of microscopic particles.

However, many fundamental and specific questions have been only partially explored to date. One of the most prominent examples is the  conflict between measurement and the laws of thermodynamics \cite{Quantum measurements and equilibration: the emergence of objective reality via entropy maximisation,On the question of measurement in quantum mechanics,Ideal projective measurements have infinite resource costs}. Can it be resolved by considering dynamics of correlations accompanying measurement as part of the equilibration process \cite{Quantum measurements and equilibration: the emergence of objective reality via entropy maximisation}?  Another important issue is the role of the initial entanglement between the system and its environment. In most studies, it is assumed that they are uncorrelated. At the same time it is known that introducing initial entanglement between the system and the environment can have a significant impact on the dynamics of the system \cite{Improving the understanding of the dynamics of open quantum systems} and on the change in its entropy \cite{Entropy dynamics in the system of interacting qubits,Entanglement versus Stosszahlansatz: Disappearance of the thermodynamic arrow in a high-correlation environment}. Exploring the role of correlations in entropy production within the context of non-Markovian processes is also appealing. For example, the Refs.~\cite{Entropy production and correlations in a controlled non-Markovian setting, Entropy Production in Non-Markovian Collision Models} raise the question: are the same correlations responsible for the negative entropy production as for non-Markovianity? The necessity of correlations between system and environment for heat transfer and, as a consequence, for the second law of thermodynamics, is demonstrated in Ref.~\cite{Correlations in quantum thermodynamics: Heat work and entropy production}. However, the role of different types of correlations in entropy dynamics is still not fully investigated. Most often, when discussing the influence of correlations on entropy change,  it is the correlations between the system and the environment that are presumed. However, are they the only ones that matter? It turns out that in some important cases, the predominant role in entropy production belongs to correlations within the environment \cite{Entropy production in open systems,Entropy production and the role of correlations in quantum Brownian motion}.

  In this work, we will focus on examining entanglement within the system. Does it influence the growth of entropy? Does it promote or hinder it? And how significant is its contribution? These questions are closely related to the dynamics of correlations. The flow of correlations was studied, for example, in Refs.~\cite{Flow of quantum correlations, Distributed correlations and information flows, Monogamy and backflow of mutual information, Overview on the phenomenon of two-qubit entanglement revivals, Dynamics of quantum entanglement, Dynamics of entanglement transfer through multipartite dissipative systems, Global correlation and local information flows, Lewis-Swan}. In Refs.~\cite{Distributed correlations and information flows} and \cite{Flow of quantum correlations}, authors consider two distinct models, where the system consists of two qubits, one of which interacts with the environment. Paper \cite{Distributed correlations and information flows} explores the mutual dynamics of two- and three-partite correlations (considering two qubits of the system and environment as three parts), demonstrating how different forms of information transform into each other. In Ref.~\cite{Flow of quantum correlations}, it is shown that the entanglement between the system and the environment is determined by the type of quantum channel and the initial entanglement within the system. The phase damping channel is responsible for the three-partite entanglement of GHZ-type, while the amplitude damping channel is responsible for W-type. For both channels, it is clearly evident that the initial entanglement between qubits contributes to the growth of correlations between the system and the environment. The mutual dynamics of the averaged entanglement of formation and von Neumann entropy of the system under the influence of several different quantum channels is studied in Ref.~\cite{Dynamics of quantum entanglement}. The averaging of entanglement of formation is separately performed for a set of pure maximally entangled states and for a set of separable ones. It is shown that in general, a decrease in entanglement is accompanied by an increase in mixedness. Comparing the graphs, it can be concluded that initial entanglement promotes an increase in entropy. 
  
A hint that internal entanglement contributes to entropy growth can also be observed from Refs.~\cite{Decoherence of two-qubit systems: a random matrix description.,Protecting quantum systems from decoherence with unitary operations}. In Ref.~\cite{Protecting quantum systems from decoherence with unitary operations}, it is demonstrated that the stronger the qubit undergoing decoherence channel is entangled with the rest of the system, the less effectively the initial multi-qubit state can be restored using the unitary operations method. In Ref.~\cite{Decoherence of two-qubit systems: a random matrix description.} authors explore the impact of entanglement within the system on decoherence. They consider two two-qubit models, where the initial states of the system are states of a specific type, and both the environment and the interaction of qubits with the environment are determined via random matrices. It turns out that in both models the purity of Bell states decays faster than the purity of product states. However, in time reversal invariance conserving model purity turns out to be a non-monotonic function of the parameter responsible for entanglement. Reduction in the purity is connected with entropy growth and for one qubit their relationship is unambiguous.

However, the question of the influence of internal entanglement has not been separately and comprehensively investigated. For example, in the majority of the aforementioned works, only systems of two qubits were considered. But will this pattern hold for larger systems? Will it become more pronounced? Is there one universal regularity? Or we can distinguish different cases with different patterns? What factors are responsible for shaping these regularities? Below, we will address these questions. We will see that due to the phenomenon of measure concentration, for Haar-random states, internal entanglement contributes to entropy growth, the slope of the fitted line increases with the size of the system, but never exceeds $29.25^\circ$. On the other hand, we will ascertain that imposing certain conditions on the initial states sample and choosing the type of interaction, can change the dependence to the opposite. This investigation can be useful in understanding the role of entanglement in the second law of thermodynamics as well as for other issues in the fields of entanglement theory and open quantum systems.

\section{Model.} 

In this paper, we consider the system consisting of a few distinguishable qubits that are not interacting between themselves. Initially, the system is in a pure state, while the degree of entanglement between the qubits is not specified. The environment is represented by a quantum harmonic oscillator. Here, we choose an infinite-dimensional environment to make the dynamics irreversible which will allow us to focus our attention on the final states. Thus, in this work, we investigate the dependence of the entropy of final states on the initial internal entanglement. We rely on the model described in detail in Ref.~\cite{Lidar}, extending their setup to more than one qubit.

We start from the well-known law of evolution of the density matrix for an open system
	\begin{equation}
		\rho_S(t)=\mathrm{Tr}_E(U \ket{\psi (0)}\bra{\psi(0)}\otimes \rho_E(0)U^\dagger),
	\label{eq:ref1}
	\end{equation}
where $\ket{\psi (0)}$ is the initial state of the system. We assume that the environment is in the Gibbs state
	\begin{equation}
		\rho_E(0)
		=\sum_{\nu=0}^{\infty}\lambda_\nu\ket{\nu}\bra{\nu}
		=\sum_{\nu=0}^{\infty}\frac {e^{-\beta E_{\nu}}} {Z} \ket{\nu}\bra{\nu}.
	\label{eq:ref2}
	\end{equation}
Where $\beta$ is the inverse temperature (later we will see that its precise value doesn't matter for our investigation). Hamiltonian of a quantum harmonic oscillator is
	$H_E=\sum_{\nu=0}^{\infty}\ E_\nu\ket{\nu}\bra{\nu}$, 
where
	 $E_\nu=\omega(\nu+1/2)$.
In this work, we study time-independent interaction Hamiltonian of the form:
	\begin{equation}
		H_{SE}
		=(A_1\otimes I_2\otimes \ldots \otimes I_2+\ldots+ I_2\otimes \ldots \otimes I_2\otimes A_n) \otimes \widehat{n}_E
		=A_S\otimes \widehat{n}_E,
	\label{eq:ref3}
	\end{equation}
where $\widehat{n}_E$ is the number operator, $\widehat{n}_E \ket{\nu}=\nu\ket{\nu}$, $A_i$ -- operators describing the interaction of $i$-th qubit with the environment, $A_S$ -- describes the interaction of the system as a whole. The interaction Hamiltonian of the form $A_S\otimes \widehat{n}_E$ for one-qubit system is used in Ref.~\cite{Lidar}, $A_S$ of such structure is used in Ref.~\cite{Monogamy and backflow of mutual information} for a system consisting of two parts. Such a form of $A_S$ implies that the qubits interact with the environment without interacting with each other.

Further, for simplicity, suppose that we can neglect the Hamiltonian of the system itself. This is justified if the interaction strength is much larger than other parameters of the system. Such a choice together with commutation of $H_{SE}$ and $H_E$ will allow us to focus on the eigenvectors of the interaction Hamiltonian. Therefore the evolution operator is\footnote{in this paper $\hbar=1$}:
	\begin{equation}
		U=e^{-itA_S\otimes \widehat{n}_E}
		e^{-itI_S\otimes\sum_{\nu=0}^{\infty}
		E_{\nu}\ket{\nu}\bra{\nu}}
		=\sum_{\nu=0}^{\infty}
		e^{-	itA_S \nu}
		e^{-itE_{\nu}}\ket{\nu}\bra{\nu}.
	\label{eq:ref4}
	\end{equation}
Then Eq.(\ref{eq:ref1}) takes the form:
	\begin{equation}
		\rho_S(t)=\mathrm{Tr}_E\bigg[\sum_{\nu=0}^{\infty} \lambda_\nu 
		e^{-itA_S\nu} \ket{\psi (0)}\bra{\psi(0)} \otimes \ket{\nu}\bra{\nu}e^{itA_S\nu}\bigg].
	\label{eq:ref5}
	\end{equation}
Let $a_j$ and  $\ket{a_j}$ denote eigenvalues and eigenvectors of $A_S$. Now we can represent $\ket{\psi (0)}$ in the eigenbasis of the operator $A_S$: $\ket{\psi (0)}=\sum_{j}c_j \ket{a_j}$, and thereby get rid of operator $A_S$ in the exponent in (\ref{eq:ref5}), replacing it with the corresponding eigenvalues. After tracing over the environment we have:
	\begin{equation}
		\rho_S(t)=\sum_{\nu,j,k}\lambda_{\nu}
		c_j c_k^*\ket{a_j}
		\bra{a_k}e^{it\nu(a_k-a_j)}.
	\label{eq:ref7}
	\end{equation} 
	
From (\ref{eq:ref7}) we see that the terms with $a_j=a_k$ do not change with time for any temperature of the environment. As for terms with $a_j\ne a_k$, it is possible to show (if we replace summation by integration and introduce high-mode cutoff of $\nu_c$), that they tend to 0 with time~\cite{Lidar}. And although the system's dynamics will depend on $\beta$, its value will not affect the final state (for greater clarity, see Eq.~(253c) in the Ref.~\cite{Lidar}). As a result of interaction with the environment, the system will lose the coherence between its pointer states, evolving into their mixture~\cite{Zurek}. Ultimately, its density matrix in the $A_S$ eigenbasis, which is the pointer basis for the chosen model, will take a diagonal form if all qubits interact and block-diagonal if some are isolated, i.e., decohere~\cite{Lidar,Quantum decoherence,Qubit disentanglement and decoherence via dephasing,Phonon decoherence of quantum entanglement: Robust and fragile states.}.

\section{Two-qubits system and different types of interactions} 

Let’s consider a two-qubit system and represent $A_S$ using the decomposition into Pauli matrices
	\begin{equation}
		A_S=
		\underbrace{
			\begin{pmatrix}
				\alpha_0+\alpha_z&\alpha_x-i\alpha_y\\
				\alpha_x+\alpha_y&\alpha_0-\alpha_z
			\end{pmatrix}}
		_{A_1}\otimes I_2
			+I_2\otimes
				\underbrace{
					\begin{pmatrix}
						\beta_0+\beta_z&\beta_x-i\beta_y\\
						\beta_x+\beta_y&\beta_0-\beta_z
					\end{pmatrix}}
				_{A_2}.
	\label{eq:ref31}
	\end{equation} 
Eigenvectors of $A_S$ and corresponding eigenvalues can be represented as:
	\begin{equation}
	\begin{matrix}
			\ket{u_p}\otimes \ket{v_p}, \ \  &\alpha_p+\beta_p,\\
			\ket{u_p} \otimes \ket{v_m},\  &\  \alpha_p+\beta_m,\\
			\ket{u_m}\otimes \ket{v_p},\  &\ \alpha_m+\beta_p,\\
			\ket{u_m}\otimes \ket{v_m}, &\ \ \alpha_m+\beta_m,
			\end{matrix}
		\label{eq:ref36}
		\end{equation}	
where $\ket{u_{p,m}}$ are eigenvectors of $A_1$, $\ket{v_{p,m}} $ - of $A_2$, $\alpha_{p,m} =\alpha_0\pm\sqrt{\alpha_x^2+\alpha_y^2+\alpha_z^2}$, $\beta_{p,m} =\beta_0\pm\sqrt{\beta_x^2+\beta_y^2+\beta_z^2}$ - eigevalues of $A_1, A_2$ correspondingly. If one qubit is isolated there is only one term in (\ref{eq:ref31}), therefore (\ref{eq:ref36}) implies two different eigenvalues. In the case when both qubits interact and the condition 
\begin{equation}
		\sqrt{\alpha_x^2+\alpha_y^2+\alpha_z^2} \ne \sqrt{\beta_x^2+\beta_y^2+\beta_z^2}
	\label{eq:refne}
	\end{equation} 
is fulfilled Eq.~(\ref{eq:ref36}) implies four different eigenvalues. If (\ref{eq:refne}) is not fulfilled there are three different eigenvalues, but we will omit such cases in this paper.

We expand the initial state vector in the basis (\ref{eq:ref36}):
	\begin{equation}
		\ket{\psi(0)}=\sum_{j,k \in \{p,m\}}c_{jk}
		\ket{u_j}\ket{v_k}.
	\label{eq:ref37}
	\end{equation} 
In the case of four different eigenvalues, the final state density matrix has the form:
	\begin{equation}
		\rho_S(t=\infty)=\sum_{j,k \in \{p,m\}}|c_{jk}|^2
\ket{u_j }\bra{u_j}
 \otimes 
\ket{v_k }\bra{v_k}.
	\label{eq:ref38}
	\end{equation}
In this paper, we work with von Neumann entropy in base 2, which in the case of pure initial states can be considered as ``the amount of decoherence introduced into the system'' \cite{Quantum decoherence}. The von Neumann entropy of the final state has the form (eigenvalues of the state (\ref{eq:ref38}) are $\theta_{jk}=|c_{jk}|^2$):
\begin{equation}
		S_N(\infty)=-\sum_{j,k \in \{p,m\}}|c_{jk}|^2\   \log_2 \ |c_{jk}|^2.
	\label{eq:ref39a}
	\end{equation} 

The case when one of two qubits remains isolated (without loss of generality, the second one) is also worth attention. Expression (\ref{eq:ref7})  implies that the existence of only two different eigenvalues leads to a combination of eigenvectors corresponding to the same eigenvalues and the final density matrix takes the form
	\begin{equation}
		\begin{matrix}
			\rho_S (\infty)
				=(c_{p_0}\ket{u_{p}} \ket{0}+c_{p_1}\ket{u_{p}} \ket{1})(c_{p_0}^*\bra{u_{p}} \bra{0}+c_{p_1}^*\bra{u_{p}} \bra{1})+\\
				\ \ \ \ \ \ \ \ \ \ \ \ \ (c_{m_0}\ket{u_{m}} \ket{0}+c_{m_1}\ket{u_{m}} \ket{1})(c_{m_0}^*\bra{u_{m}} \bra{0}+c_{m_1}^*\bra{u_{p}} \bra{1})
			\end{matrix},
	\label{eq:ref13}
	\end{equation} 
where we use expansion of the initial state vector
 \begin{equation}
 \ket{\psi(0)}=
 c_{p_0}\ket{u_{p}} \ket{0}+c_{m_0}\ket{u_{m}} \ket{0}+c_{p_1}\ket{u_{p}} \ket{1}+c_{m_1}\ket{u_{m}} \ket{1}.
 \label{exp2}
 \end{equation}
The eigenvalues of $\rho_S (\infty)$ are
\begin{equation}
	\theta_{p}=|c_{p1}|^2+|c_{p2}|^2, \theta_{m}=|c_{m1}|^2+|c_{m2}|^2.	
	\label{eq:ref11b}
	\end{equation} 
Corresponding entropy
	\begin{equation}
		S_{N}(\infty) =-\sum_{j=p,m}(|c_{j1}|^2+|c_{j2}|^2)\  log_2\ (|c_{j1}|^2+|c_{j2}|^2).
	\label{eq:ref16}
	\end{equation} 
Note that from expressions (\ref{eq:ref39a}) and (\ref{eq:ref16}), it is evident that the entropy of the final states does not depend on the specific values of $\alpha_i,\beta_i$ in (\ref{eq:ref31}), although they do affect the eigenvalues (\ref{eq:ref36}) and consequently the system state evolution (\ref{eq:ref7}).

\begin{figure}[h]
\begin{minipage}[h]{0.96\linewidth}
\center{\includegraphics[width=1\linewidth]{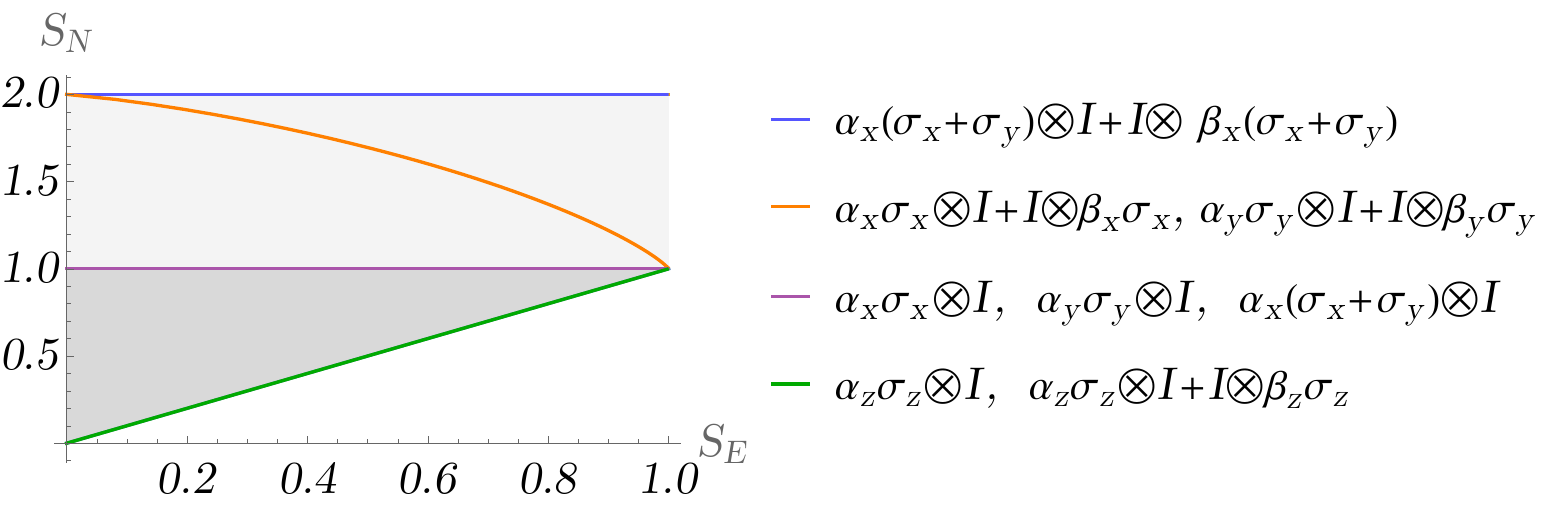}} \\
\end{minipage}

\caption{Final entropy vs initial entanglement entropy for states of type (\ref{eq:ref20})  for some choices of $A_S$ (lines of different colors correspond to different interactions). The values of the coefficients $\alpha_{x},\alpha_{y},\alpha_{z},\beta_{x},\beta_{y},\beta_{z}$ do not affect the final result; what matters is that condition (\ref{eq:refne}) is satisfied (if (\ref{eq:refne}) is not satisfied (\ref{eq:ref36}) implies three different eigenvalues and the picture will differ). The monotonicity of the curves is determined by the form of $|c_{jk}|^2$ (in the cases when one and two qubits interact these are coefficients in expansions (\ref{eq:ref37}) and (\ref{exp2}) correspondingly). The allowed area (for any possible initial states and interactions of type (\ref{eq:ref31})) is shaded in grey (see Appendix A). The area allowed in the case when only one qubit interacts is shaded in a darker grey.}

\label{Fig.1}
\end{figure}

Since, in this section, we consider a system of two qubits in a pure state, we choose the entanglement entropy as a measure of entanglement. It is easy to see that there is no simple connection between the entanglement entropy of the initial state and the final entropy (\ref{eq:ref39a}) (the same for (\ref{eq:ref16})). Figure~1 depicts various dependencies of $S_{N}(\infty)$ on the entanglement of initial states of type:

\begin{equation}
		\gamma_1\ket{00}+\gamma_2\ket{11}.
	\label{eq:ref20}
	\end{equation} 
Thus, we observe that even for a specific type of states, different choices of interaction can yield opposite dependencies of $S_{N}(\infty)$ on initial entanglement. At the same time, we cannot associate the type of dependency only with the form of interaction, because $|c_{jk}|^2$ (in (\ref{eq:ref39a}) and (\ref{eq:ref16})) depends on both:  interaction and initial state.

The most interesting is the interplay between the Schmidt vectors and the eigenbasis of the operator $A_S$. According to the Schmidt theorem, if we have a pure state describing a system of two subsystems, then we can always choose bases for each of the subsystems in such a way that the state will be represented through the Schmidt coefficients. If for some initial state, the Schmidt basis coincides with the set of eigenvectors (\ref{eq:ref36}) of the interaction operator $A_S$, then, after the interaction, all off-diagonal elements of the matrix in the Schmidt basis become zero (regardless of whether one or two qubits interact with the environment), and non-zero eigenvalues of such a matrix are equal to the Schmidt coefficients. Thus, the entropy of the final state is equal to the entropy of the entanglement of the initial one. This is precisely what we see for states (\ref{eq:ref20}) and interactions
 \begin{equation}
		A_S=\alpha_z \sigma_z\otimes I,
		A_S=\alpha_z \sigma_z\otimes I+I\otimes \beta_z \sigma_z,
			\label{eq:ref21}
	\end{equation} 
where in the second expression $\alpha_z, \beta_z$ satisfy (\ref{eq:refne}). The same is true for states $\gamma_1(\ket{00}+\ket{11})+\gamma_2(\ket{10}+\ket{01})$ and interactions $A_S=\alpha_x \sigma_x\otimes I$, 
 $A_S=\alpha_x \sigma_x\otimes I+I\otimes \beta_x \sigma_x $ since they can be obtained from (\ref{eq:ref20}) and (\ref{eq:ref21}) as a result of transformations of qubit bases. 
 
 In the general case, when the Schmidt basis can differ from the eigenvector basis (\ref{eq:ref36}), the final entropy will always be greater than or equal to the initial entanglement entropy (for the proof see Appendix A\footnote{We provide proof only for cases of two and four different eigenvalues, as for three different eigenvalues, expression (\ref{S_NS_E(0)}) may not hold. For instance, by straightforward calculation, one can verify that the interaction of type $\alpha_z( \sigma_z\otimes I+I\otimes \sigma_z)$ does not alter states of type $\gamma_1\ket{01}+\gamma_2\ket{10}$. Thus, $S_N(\infty)=0$ despite for all states (except $\ket{01}$ and $\ket{10}$) $S_E(0))> 0$. At the same time, expression (\ref{eq:ref61}) will hold for this example. The problem of robust and fragile entangled states is discussed in Ref.~\cite{Phonon decoherence of quantum entanglement: Robust and fragile states.}, as the example the authors consider coupling operator with three different eigenvalues.}): 
  \begin{equation}
  S_N(\infty)\geq S_E(0).
  \label{S_NS_E(0)}
  \end{equation}  
This condition determines the allowed area on the plane $S_N(\infty)S_E(0)$ and applies to any pure initial states and interactions of the form (\ref{eq:ref31}). For the case when one qubit interacts, the allowed region in Figure~1 is shaded in dark gray, while for the case when both qubits interact, the allowed region includes both light gray and dark gray parts. Thus, if one of the qubits remains isolated, the final entropy takes smaller or equal values (see Appendix A). The reason is that when both qubits of the system interact with the environment the final matrix in the basis of eigenvectors of the interaction operator takes a diagonal form. If one qubit is isolated, some of the off-diagonal elements remain non-zero. The issue of the density matrix change over time when one and both qubits interact with the environment is discussed in Ref.~\cite{Qubit disentanglement and decoherence via dephasing}, where the authors compare the rate of disentanglement and the rate of decoherence of each individual qubit. However, they are not interested in the entropy of states and do not discuss the influence of the initial entanglement on its growth.

 In the chosen model for pure initial states, it can be said that the increase in entropy between initial and final states is always greater than or equal to the decrease in the entanglement of the system: 
   \begin{equation}
		S_N(\infty)-S_N(0)\geq S_E(0)-S_E(\infty).
		\label{eq:ref61}
	\end{equation} 
Here $S_E(\infty)=0$ because the final states (\ref{eq:ref38}) and (\ref{eq:ref13})\footnote{State (\ref{eq:ref13}) can be represented in the form that makes its disentanglement more clear $\rho_{S} (\infty) 
		=\sum_{j=p,m}
		\bigr(\ket{u_j}\bra{u_j})
		\otimes
			\begin{pmatrix}
				c_{j1}\\
				c_{j2}
			\end{pmatrix}
			\begin{pmatrix}
				c_{j1}^*&c_{j2}^*
			\end{pmatrix}$.}
 are disentangled. Thus, in the model under consideration, the internal entanglement entirely transforms into the system-environment correlation, and classical correlation within the system (since in general, (\ref{eq:ref38}) and (\ref{eq:ref13}) are not direct products). Therefore, the results we see in Figure~1 do not contradict the property of quantum monogamy \cite{Multipartite Entanglement Measure and Complete Monogamy Relation}, according to which qubits maximally entangled with each other cannot be simultaneously entangled with the environment.

\section{Random states for 2, 3, 4, 5, and 6 qubits}

Although the entanglement of the initial states imposes certain constraints on the potential values of the final entropy, the allowed region in Figure~1 is quite extensive. As the system grows and the number of qubits interacting with the environment increases, the maximum possible values of the final entropy become increasingly larger, while the states of type $(\ket{0\dots0}+\ket{1\dots1})/\sqrt{n}$ in the result of $\sigma_z$-interaction give $S_N(\infty)=1$. Therefore it is clear that the proportion of the forbidden region will decrease. However, this does not mean that we cannot identify any patterns in the relationship between the initial entanglement and the final entropy. We investigate the situation statistically. For these purposes, we generate a set of Haar-random initial pure states with the help of Hurwitz parametrization \cite{Composed ensembles of random unitary matrices} (see Appendix~B). If we consider an individual qubit, the generated states will be uniformly distributed in all directions originating from the center of the Bloch sphere. Therefore, all $\sigma_{x,y,z}$ are treated symmetrically. For each of the initial states, we determine the measure of entanglement and the entropy of the corresponding final state.

Since in this section, we also want to consider systems consisting of three or more qubits, we cannot continue to work with the entanglement entropy. As a measure of entanglement, we choose global entanglement or the Meyer-Wallach measure, which was first introduced in Ref.~\cite{Wallach-Meyer}. Subsequently, it was reformulated in a simpler form that makes clear its interpretation as the average purity of the system's qubits \cite{An observable measure of entanglement for pure states of multi-qubit systems}. In  \cite{Wallach-Meyer, An observable measure of entanglement for pure states of multi-qubit systems} the measure was normalized in such a way that its maximum value is unity for any number of qubits. However, in our investigation, we would like to observe that a larger system can possess greater entanglement. Hence, we rewrite global entanglement in the following form (the corresponding form appeared in Ref.~\cite{Multipartite Entanglement Measure and Complete Monogamy Relation} for a system of three qubits and was referred to as a tripartite tangle):
		 \begin{equation}
		Q(\rho_S)=n-\sum_{i=1}^{n}Tr\rho_i^2,
		\label{eq:ref50}
	\end{equation} 
where $\rho_i$ - density matrix of $i$-th qubit is obtained from $\rho_S$ by tracing over all other qubits. Further, we will simply refer to $Q$ as entanglement.

Expression for final entropy (\ref{eq:ref39a}) can be generalized to the case of $n$ qubits. If all eigenvalues of $A_S$ are different:
\begin{equation}
		S_N(\infty)
		=-\sum_{j_1\ldots j_n}|c_{j_1 \ldots j_n}|^2\  log_2 \ |c_{j_1 \ldots j_n}|^2,
	\label{eq:ref39ac}
	\end{equation} 
where $c_{j_1 \ldots j_n}$ are coefficients in the expansion of the initial state vector in terms of the basis of eigenvectors of the operator $A_S$:
 
\begin{equation}
		\ket{\psi(0)}=\sum_{j_1=p,m \ldots j_n=p,m}
		c_{j_1 \ldots j_n}
		\ket{u_{j_1}}\otimes \ldots \otimes\ket{u_{j_n}}.
			\label{eq:ref37q}
	\end{equation} 
Here $\ket{u_{j_i}}$ are eigenvectors of operators $A_i$ in (\ref{eq:ref3}).  If some of the qubits are isolated, then, similarly to how it was done in (\ref{eq:ref13}), it is necessary to sum all $|c_{j_1 \ldots j_n}|^2$ corresponding to equal eigenvalues of $A_S$.

\begin{figure}[h]
\begin{minipage}[h]{0.48\linewidth}
\center{\includegraphics[width=1\linewidth]{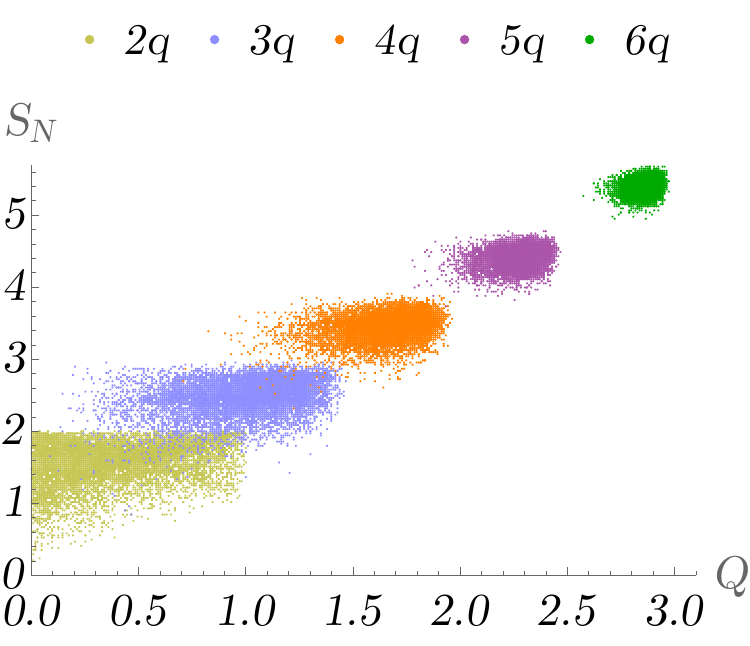}} a) \\
\end{minipage}
\hfill
\begin{minipage}[h]{0.48\linewidth}
\center{\includegraphics[width=1\linewidth]{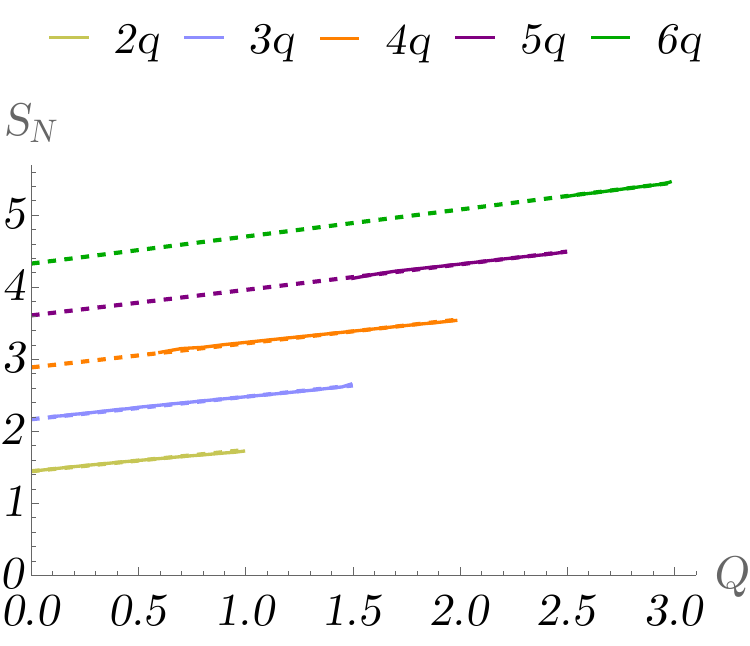}} b)  \\
\end{minipage}

\caption{Final entropy vs initial entanglement for systems of different sizes. a) Results of generating $10^4$  random states (for each system size). b) Approximate average lines (solid) align with approximating straight lines (dashed). They indicate that, on average, the initial entanglement contributes to the growth of entropy. The clustering around mean values in a) and the increase in the approximating lines in b) are based on the phenomenon of measure concentration.}

\label{Fig.2}
\end{figure}

Figure~2a) shows the generation results for 2-6-qubit systems (in Figures systems consisting of $n$ qubits we denote by $nq$). As the system grows, the points increasingly cluster around the average values of the initial entanglement and final entropy. Such behavior is very well-known under the name of concentration of measure or Levy's lemma \cite{Levys lemma,Aspects of generic entanglement}. 

 The average dependencies are depicted in Figure~2b) with solid lines. To obtain them, we generated $10^7$ Haar-random states and averaged  $S_N$ over certain intervals $\Delta Q$ for which we collected sufficient statistics. These average dependencies exhibit nearly linear behavior and align with the dashed lines. The dashed lines represent straight lines passing through the mean points for all Haar-random states and mean points for separable random states generated separately (using the method described in Appendix~B). We will refer to them as approximating straight lines.

For greater accuracy and further convenience let's write down the mean values of initial entanglement and final entropy, their variations, and average $S_N$ at the extreme points. We obtain $\overline{S}_N,\overline{Q}$ by averaging over all generated Haar-random states, $\overline{S}_N(Q=0)$ - by averaging over random separable states generated separately, $\overline{S}_N(Q_{max})$ - using approximating straight lines. We round variations to three decimal places and all other quantities to two decimal places: 
\begin{equation}
			\begin{matrix}
				\ \ \ \ \ &2q&3q&4q&5q&6q\\
				\hline
				\overline{S}_N(Q=0)&1.44&2.16&2.89&3.61&4.33\\
				\hline
				\overline{S}_N(Q_{max})&1.74&2.63&3.55&4.49&5.45\\
				\hline
				\overline{S}_N&1.56&2.48&3.43&4.41&5.40\\
				\hline
				\overline{Q}&0.40&1.00&1.65&2.27&2.86\\
				\hline
				Var(S_N)&0.076&0.053&0.031&0.017&0.009\\
				\hline
				Var(Q)&0.068&0.054&0.024&0.008&0.002.
			\end{matrix}	
			\label{eq:ref600}
	\end{equation} 
	
Notice that for systems with $Q=0$, the average final entropy, up to numerical errors, is proportional to the number of qubits in the system: $0.72n$. This is not surprising, since initial states for n-qubit system, represented by direct products 
$ \big(\sum_{j=p,m}c_{j_1}\ket{u_{j_1}}\big)\otimes \dots \otimes (\sum_{j=p,m}c_{j_n}\ket{u_{j_n}})$ will ultimately evolve into direct products 
$\big(\sum_{j=p,m}|c_{j_1}|^2\ket{u_{j_1}}\bra{u_{j_1}}\big)\otimes \dots \otimes \big(\sum_{j=p,m}|c_{p_n}|^2\ket{u_{j_n}}\bra{u_{j_n}}\big)$.	
	
At the same time, for larger $Q$, the proportionality law does not hold due to two factors. On the one hand, the final states contain classical correlations\footnote{By analogy with (\ref{eq:ref38}), final states can be expressed as
$		\sum_{j_1\dots j_n \in \{p,m\}}|c_{j_1\dots j_n}|^2
\ket{u_{j_1} }\bra{u_{j_1}}
 \otimes 
 \dots
  \otimes 
\ket{u_{j_n} }\bra{u_{j_n}}$, which in general are not direct products.}, which lead to non-additivity of entropy, making its values smaller. On the other hand, the average initial and consequently final entropies of a single qubit increase with the system size: 0.48, 0.73, 0.87, 0.93, 0.97 for the initial and 0.84, 0.92, 0.96, 0.98, 0.99 for the final states for systems consisting of 2, 3, 4, 5 and 6 qubits correspondingly. This growth is associated with the phenomenon of measure concentration. The mentioned two factors result in the obtained values of $\overline{S}_N$. These values are bigger than $\overline{S}_N(Q=0)$ and we observe a positive slope of the approximating average lines.

Moreover, as the system grows, variances decrease, and $\overline{S}_N$ and $\overline{Q}$ increasingly approach their maximal possible values $S_{Nmax}=n$ and $Q_{max}=n/2$. For 1, 2, 3, 4, 5, 6-qubit systems quantity $\overline{S}_N/S_{Nmax}$ takes values 0.72, 0.78, 0.83, 0.86, 0.88, 0.90. Quantity $\overline{Q}/Q_{max}$  for 2, 3, 4, 5, 6-qubit systems takes values 0.40, 0.67, 0.82, 0.91,  0.95. From Figure~2b), it is evident that the slope angle of the approximating lines increases.  For 2, 3, 4, 5, and 6 qubits, it is $16^\circ, 17^\circ, 18^\circ,20^\circ, 21^\circ$  (here we consider $S_N$ and $Q$ in the same scale and round angles to the nearest whole numbers). But even for the infinite number of qubits, it will not exceed $29.25^\circ$, since the tangle of maximal angle is $(S_{Nmax}-S_N(Q=0))/Q_{max}=0.56$. 

The tendency of $\overline{Q}, \overline{S}_N$ to approach the maximum possible values originates from the measure concentration phenomenon. The dependence of the average entanglement on the system size was discussed in Ref.~\cite{Average entropy of a subsystem,{Quantum entanglement in random physical states}} for large systems. 
The tendency of $\overline{S}_N$ to approach $S_{Nmax}$ can be explained as follows. Consider some $A_S$-interaction and represent initial states in the eigenbasis of the corresponding operator. The predominant fraction of random states consists of states with values of $|c_{j_1 \ldots j_n}|^2$-s relatively close to each other. Due to the measure concentration phenomenon, as the system grows, the spread of $|c_{j_1 \ldots j_n}|^2$-s becomes increasingly smaller. At the same time, according to (\ref{eq:ref39ac}), states with the same $|c_{j_1 \ldots j_n}|^2$-s give the maximum possible values of $S_N$. Therefore measure concentration is responsible for the growth of average lines in Figure~2b) and for the increase of their slope.

The effect of entanglement on entropy growth is noticeably smaller than the effect of the size of the system. We can estimate the fraction of entropy growth associated with entanglement in the overall entropy change using the expression:
\begin{equation}
			(\overline{S}_N-\overline{S}_N(Q=0))/\overline{S}_N,
			\label{eq:refSNQ}
	\end{equation} 
For 2, 3, 4, 5, 6-qubit systems (\ref{eq:refSNQ}) gives 0.08, 0.13, 0.16, 0.18, 0.20 correspondingly. However, even for large systems, it will not exceed 0.28.

To further investigate the relationship between two quantities, we use the Pearson correlation coefficient. It varies within the range $[-1,1]$, and the closer its absolute value is to one, the stronger the linear correlation dependency. The Pearson coefficient for 2, 3, 4, 5, 6-qubit systems takes values: 0.27, 0.30, 0.28, 0.25, 0.21. Thus, starting from 4 qubits, due to the clustering of values of initial entanglement and final entropy around their means, the clarity of the linear dependence diminishes. 

 \begin{figure}[h]
\begin{minipage}[h]{0.60\linewidth}
\center{\includegraphics[width=1\linewidth]{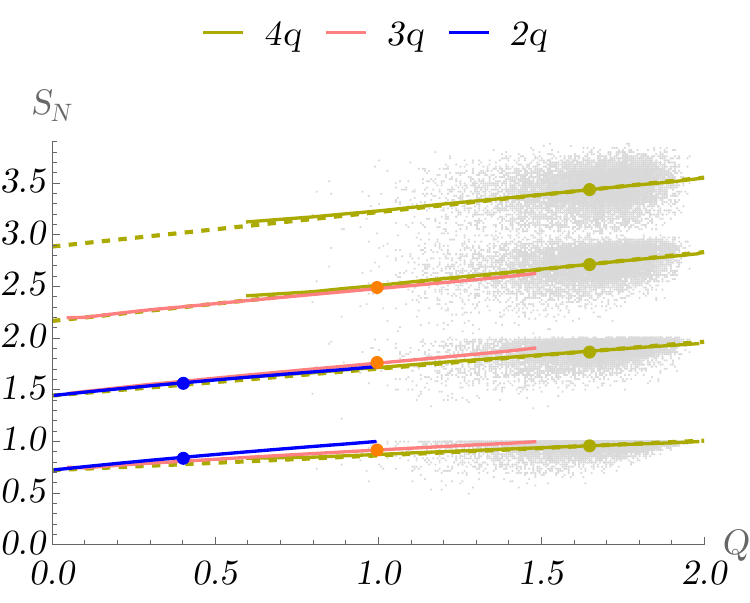}} a) \\
\end{minipage}
\hfill
\begin{minipage}[h]{0.36\linewidth}
	 \begin{equation}
		\begin{matrix}
			\begin{matrix}
				\mathrm{Number\ of}\\
				\mathrm{interacting}\\
				\mathrm{qubits}
			\end{matrix}
		&\begin{matrix}
		\mathrm{System}\\
		\mathrm{size}
		\end{matrix}
		&\overline{S}_N\\
		\hline
 		1 &2q&0.84\\
		\hline
		  1 &3q&0.92\\
		  \hline
		   1 &4q&0.96\\
		   \hline
   		 2&2q&1.56\\
		 \hline
  		   2&3q&1.76\\
		   \hline
 		     2 &4q&1.87\\
		     \hline
 		     3&3q&2.48\\
		     \hline
  		    3&4q&2.71\\
		    \hline
  		     4&4q&3.43\\
		\end{matrix}
		\label{eq:ref602}
	\end{equation} 
	\center{b)}
\end{minipage}

\caption{ The influence of the number of interacting qubits for systems of different sizes. a) Light gray dots correspond to $10^4$ random states for the 4-qubit system. The greater the number of interacting qubits, the higher the dots are arranged, and the greater their vertical blurring. The approximate average lines (solid) show dependencies for different system sizes and different numbers of interacting qubits. In regions where there was little data, we supplemented them with approximating straight lines (dashed). The lower group of lines corresponds to the case when only one qubit interacts with the environment, the second - when two, and so on. Systems of different sizes are represented by lines of different colors. Bold dots in the figure indicate mean points. b) The mean values of final entropy for different numbers of interacting qubits for systems of different sizes.}

\label{Fig.3}
\end{figure}

Now let's consider a situation where only a part of the system's qubits interact with the environment, while the rest remain isolated. Figure~3a) presents approximate average lines for a 4-qubit system. We see that the number of interacting qubits has the biggest impact on entropy growth. At the same time entanglement, on average, also contributes to an increase in entropy for any number of interacting qubits, and the greater their number, the stronger dependence: for 1, 2, 3, 4 interacting qubits of a 4-qubit system the slope is $8^\circ, 14^\circ,18^\circ, 18^\circ$ correspondingly. However, the fraction of the increase in entropy due to entanglement, compared to the overall value, decreases. For a 4-qubit system with 1, 2, 3, 4 interacting qubits (\ref{eq:refSNQ}) gives 0.25, 0.23, 0.20, 0.16 correspondingly.  It is also worth noting that for small systems the size of the entire system matters, even if the same number of qubits interact. If only one qubit interacts then for 2, 3, 4-qubit systems (\ref{eq:refSNQ}) gives 0.14, 0.21, 0.25 correspondingly. Such growth of values can be explained by the fact that the larger the system, the stronger (on average) each qubit is entangled with the rest of the system.

\section{Some special samples} 

In the previous section, we examined states uniformly distributed throughout the entire space of pure states. However, since the correlation between the initial internal entanglement and the final entropy is not perfect, it is clear that the choice of state samples plays an important role in the final dependence. In this section, we will consider three cases of special sampling: a system consisting of clusters of entangled subsystems, a system with the fixed average excited states occupations, and a non-symmetric generalization of Dicke states.

\subsection{A system consisting of several independent subsystems.}

\begin{figure}[h]
\ \ \ \begin{minipage}[h]{0.40\linewidth}
\center{ \includegraphics[width=1\linewidth]{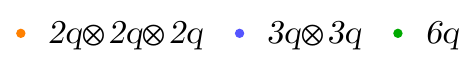}}\\
\end{minipage}
\hfill
 \ \ \ \begin{minipage}[h]{0.54\linewidth}
\center{\includegraphics[width=1\linewidth]{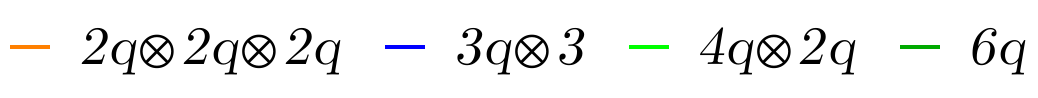}}\\
\end{minipage}
\vfill
\begin{minipage}[h]{0.48\linewidth}
\center{\includegraphics[width=1\linewidth]{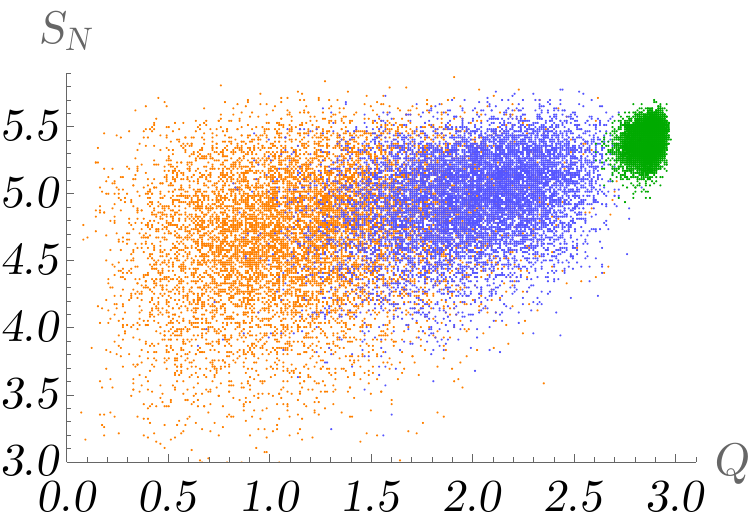}} a)\\
\end{minipage}
\hfill
\begin{minipage}[h]{0.48\linewidth}
\center{\includegraphics[width=1\linewidth]{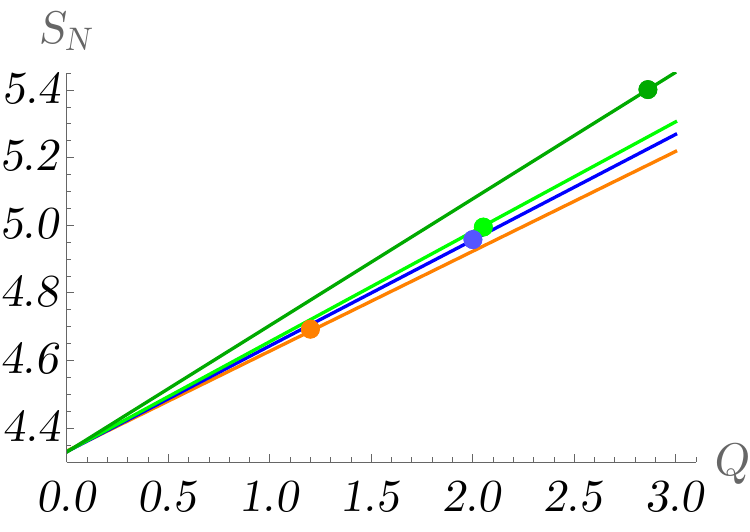}} b)\\
\end{minipage}

\caption{Six qubit system consisting of clusters of entangled subsystems with different entanglement depth. a) Final entropy vs initial entanglement for $10^4$ generated states for different entanglement depths. The larger the entanglement depth, the stronger the clustering. b) The entanglement depth of subsystems affects the mean values (bold dots in the figure indicate mean points) and the slope of approximating straight lines.}

\label{Fig.6}
\end{figure}

So far, we have been considering the entanglement effect in general, regardless of how it is distributed within the system. In this subsection, we compare the entropy growth in cases when the system consists of a few subsystems that are not correlated with each other. We represent the state of the entire system as a product state of subsystem states, each of which is Haar-random. Since the separable subset is of measure zero, by limiting the subsystem size, we set the depth of entanglement. The issue of entanglement distribution within a system and methods for theoretical and experimental determination of entanglement depth has been actively discussed in the literature for the past two decades (for review see \cite{Entanglement certification from theory to experiment}). Therefore, investigating the impact of entanglement depth on entropy growth can contribute to a more comprehensive understanding of the matter.

  It can be stated in advance that the smaller the size of subsystems, the smaller the average values of entanglement and final entropy, and the less the points will concentrate around the mean values. This is exactly what we see in Figure~4 for systems consisting of six qubits. Thus, a greater entanglement depth promotes entropy growth on average. A similar pattern can be observed for the approximating straight lines in Figure~4b). For systems consisting of independent subsystems, entropy and entanglement are simply the sums of the corresponding quantities of the subsystems. Therefore for all systems with the same entanglement depth, the approximating lines have the same slope angle. When we increase the size of subsystems, the average entropy and average entanglement tend to their maximum possible values, due to the measure concentration phenomenon. However, the ratio of convergence rates of the two quantities varies, which is more pronounced for smaller entanglement depths. For example, for systems consisting of only one subsystem, we calculate using the data from the previous section $(\overline{S}_N^{3q}-\overline{S}_N^{2n})/(\overline{Q}^{3q}-\overline{Q}^{2q})\approx1.53<(\overline{S}_N^{6q}-\overline{S}_N^{5n})/(\overline{Q}^{6q}-\overline{Q}^{5q})\approx1.68$ (the superscripts denote the size of the system). Thus, the larger entanglement depth leads to a larger slope angle of the approximating lines. This is what we observe in Figures~4b) and 2b). In conclusion, both in terms of the mean values of the final entropy and in terms of its average dependencies on the initial entanglement, the internal entanglement depth promotes the growth of von Neumann entropy.

\subsection{Dependence of the excited state occupation}

In this subsection, we demonstrate that the presence of additional constraints in the model may affect the dependence of entropy growth on internal entanglement. The hint that this might take place can be seen in the literature. For example, it is known that, for integrable and non-integrable spin chains (integrability implies the presence of additional symmetries), the average entanglement entropy between parts differs \cite{Entanglement and matrix elements of observables in interacting integrable systems}. In Ref.~\cite{Generic entanglement entropy for quantum states with symmetry} it is shown that choosing random states with particular symmetries can affect the measure concentration and, consequently, the degree of entanglement in random states. For instance, states with permutation symmetry are much less entangled than random states in general. Similarly, in Ref. \cite{Decoherence of two-qubit systems: a random matrix description.} two models with different symmetries give different dependencies of the system's purity dynamics on initial internal entanglement.

In this subsection, we consider 2 and 3-qubit systems and choose only those initial states in which all qubits have the same values of the average occupation of the excited state 
\begin{equation}
		\left\langle {E_i} \right\rangle=Tr(\rho_{i}
		\ket{1}\bra{1})=E,
		\label{eq:ref37c}
	\end{equation}
where $i$ is the number of a qubit. The quantities $\left\langle {E_i} \right\rangle$ can be interpreted as average energies of qubits before the interaction between the system and environment was turned on if we have a non-zero Hamiltonian of the system itself. This does not contradict our model if we assume that the interaction between the system and the environment is strong enough to allow us to neglect the system's internal evolution since it is significantly slower. On the other hand, if the qubits are implemented using spin-1/2 particles instead of  (\ref{eq:ref37c}) we could consider $Tr(\rho_{i} (-\ket{0}\bra{0}+\ket{1}\bra{1})/2)=E$ that would represent a constraint on the average magnetization of the qubit (although the concept of magnetization typically applies to spin systems as a whole, e.g. \cite{Entanglement and matrix elements of observables in interacting integrable systems}). In this case, in Figures~5 and 6, we would obtain similar results, but symmetric with respect to $E = 0$.

Condition (\ref{eq:ref37c}) implies that in the Bloch ball picture, all states lie in the planes $z_i=E$. Therefore, the results will be not symmetric with respect to changes in the coefficients in (\ref{eq:ref31}). Assuming that all qubits of the system interact with the environment in the same way, we now examine $A_i=\sigma_z$ (dephasing interaction without energy exchange) and  $A_i=\sigma_x$ (interaction with energy exchange) separately, while $\sigma_y$ is treated symmetrically with $\sigma_x$.

The results for 2-qubit system when both qubits interact with the environment are presented in Figure~5a),b).  To obtain this distribution we selected states satisfying (\ref{eq:ref37c}) from all random states. To accumulate sufficient statistics, we rounded values of $\left\langle {E_i} \right\rangle$ to 0.001.

 The constraint (\ref{eq:ref37c}) reduces the allowed regions compared to Figure~2a). The picture is symmetric with respect to $E=E_{max}/2=0.5$, in other words, for example, the regions for $E=0.2$ and 0.8 coincide, as well as for 0.1 and 0.9.  The farther $E$ from 0.5, the smaller the region. The position of the region and, consequently, the average values of the initial entanglement and final entropy also depend on $E$. 
The farther $E$ is from 0.5, the smaller values for $Q$ and $S_z$ (von Neumann entropy for $\sigma_z$-interaction) are allowed and the smaller are $\overline{Q}, \overline{S}_z$, whereas $S_x$ (von Neumann entropy for $\sigma_x$-interaction) is limited to larger values and has larger $\overline{S}_x$. The mean values are given in Appendix C. Observation concerning average entanglement aligns with the findings presented in Ref.~\cite{Entanglement entropy of eigenstates of quantum chaotic Hamiltonians} for a one-dimensional lattice with bosons: the closer the average site occupancy\footnote{It is worth noting that the average site occupancy in Ref.~\cite{Entanglement entropy of eigenstates of quantum chaotic Hamiltonians}  is defined as the ratio of the number of particles to the number of lattice sites, i.e. it is related, but not the same notion.}
 to $1/2$, the larger the average entanglement entropy between the parts of the system.
It is also noticeable that approximate average lines characterizing the dependencies $S_{z,x}(Q)$, unlike in the general case in Section 4, are far from being straight. Remarkably, while for $\sigma_x$, we still see that entanglement promotes an increase in entropy, for $\sigma_z$  - diminishes it. However, for $\sigma_z$, there is no contradiction with the increasing lines in Figure~2 because they were formed by states with varying energies of qubits. Thus, different parts of the approximating lines in Figure~2 correspond to initial states with different average deviations of $E$ from $0.5$. 

\begin{figure}[h]
\begin{minipage}[h]{0.44\linewidth}
\center{\includegraphics[width=1\linewidth]{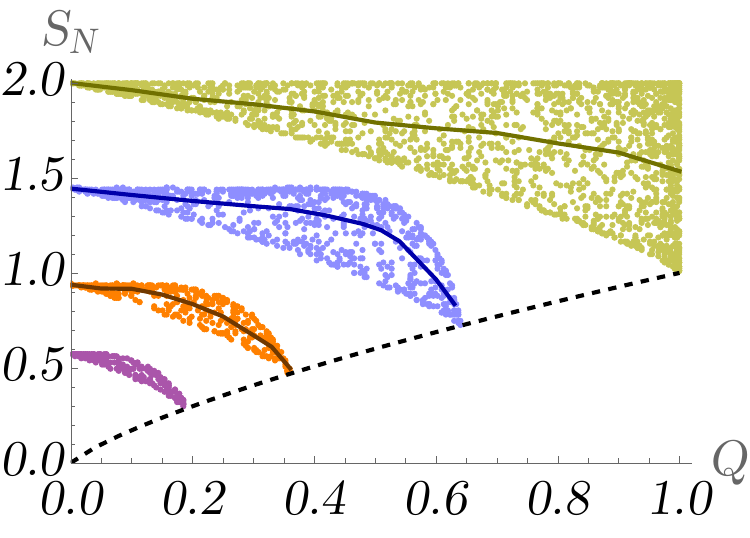}} a) $\sigma_z$-interaction \\
\end{minipage}
\hfill
\begin{minipage}[h]{0.08\linewidth}
\center{\includegraphics[width=1\linewidth]{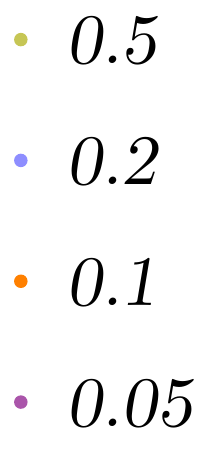}} \\
\end{minipage}
\hfill
\begin{minipage}[h]{0.44\linewidth}
\center{\includegraphics[width=1\linewidth]{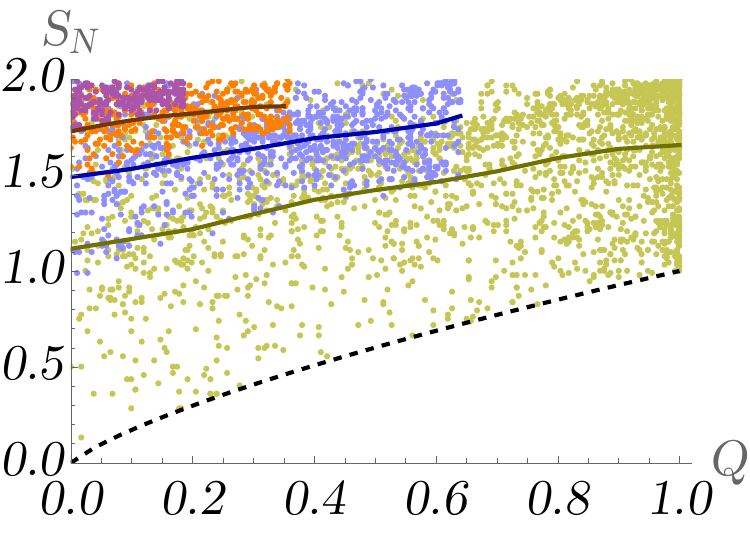}} b) $\sigma_x$-interaction\\
\end{minipage}
\vfill
\begin{minipage}[h]{0.44\linewidth}
\center{\includegraphics[width=1\linewidth]{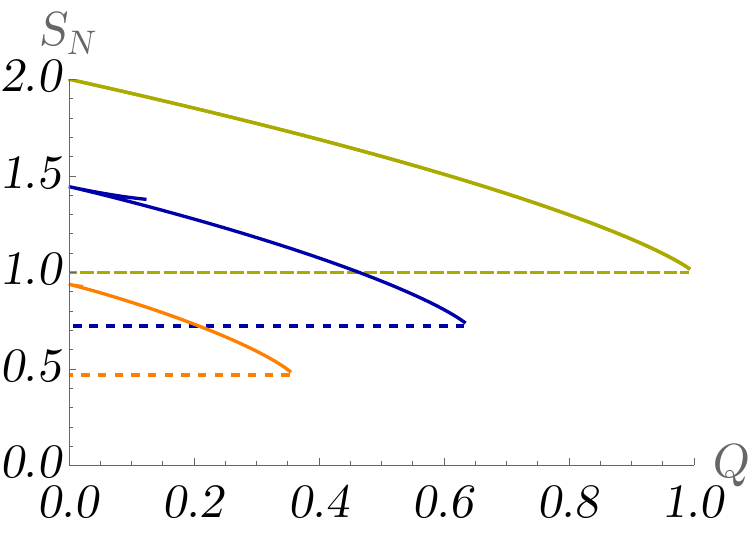}} c) $\sigma_z$-interaction\\
\end{minipage}
\hfill
\begin{minipage}[h]{0.08\linewidth}
\center{\includegraphics[width=1\linewidth]{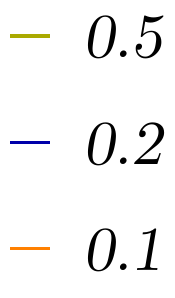}} \\
\end{minipage}
\hfill
\begin{minipage}[h]{0.44\linewidth}
\center{\includegraphics[width=1\linewidth]{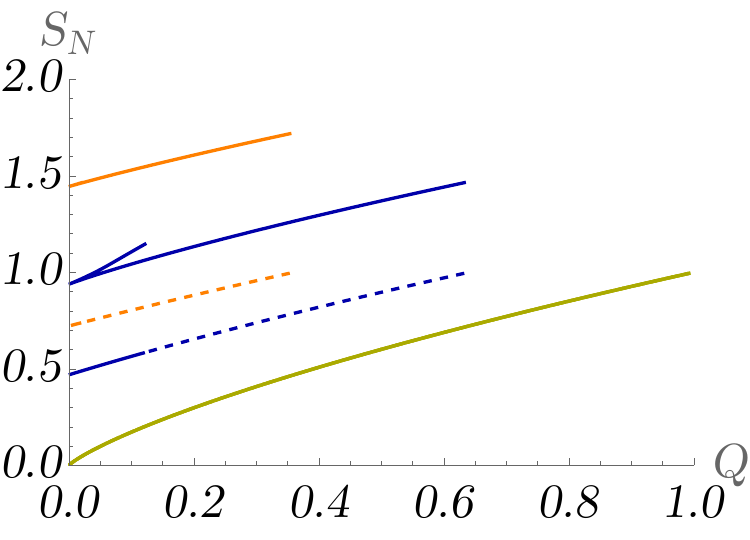}} d) $\sigma_x$-interaction\\
\end{minipage}

\caption{Imposing condition (\ref{eq:ref37c}) results in opposite dependencies of final entropy on initial entanglement for $\sigma_z$- and $\sigma_x$-interactions. Different colors correspond to different values of $E$. a),b) Regions occupied by the generated states. The dashed line shows the boundary of the region allowed for any pure states, in a) it is formed by states $\sqrt{1-E}\ket{00}+\sqrt{E}\ket{11}$, in b) by states (\ref{border}) with $E=0.5$. In a), b) approximate average lines are shown for $E=0.5, 0.2, 0.1$. c), d) States with real components having the same sign form segments of lines. In the case where both qubits interact with the environment, we draw them with solid lines; when only one qubit interacts, we use dashed lines.}

\label{Fig.5}
\end{figure}

The lower boundaries of the regions with certain values of $E$ in Figure~5a),b) are determined by states of the form
\begin{equation}
\begin{pmatrix}
			\sqrt{1-x-E}&
			\sqrt{x}&
			\sqrt{x}&
			\sqrt{E-x}
		\end{pmatrix}
		^T,
		\label{border}
\end{equation} 
where $x\in[0, \mathrm{min}\{E,1-E\}]$. Lines corresponding to (\ref{border}) are shown in Figures~5c),d) in \textit{solid}. We receive states (\ref{border}) if in addition to (\ref{eq:ref37c}) require the equality of all phases in (\ref{H1}). This is equivalent to selecting only real states with all components having the same sign. Incorporating random phases shifts the points toward higher entanglements\footnote{For states of type 
$
\begin{pmatrix}
			\sqrt{1-x-E} e^{i\varphi_1}&
			\sqrt{x}e^{i\varphi_2}&
			\sqrt{x}e^{i\varphi_3}&
			\sqrt{E-x}e^{i\varphi_4}
		\end{pmatrix}^T $  the final entanglement is $Q=1 +2 (1 - E) E +2 x( 2 x-1-2 \sqrt{1 - E - x}\sqrt{E - x}\ cos(\varphi_1 - \varphi_2 - \varphi_3 + \varphi_4))$. }.
 For $\sigma_z$-interaction random phases do not alter the final entropies but make their values larger for $\sigma_x$.

Now we can explain the curious fact that when we choose interaction operator $A_i$ with the same set of eigenvectors as the excited state occupation operator, we observe a negative dependency. Such a choice of interaction operator implies that the final state is obtained simply by setting the off-diagonal elements of the initial density matrix to zero, and for (\ref{border}) we receive: $\rho(t=\infty,E)=diag\{1-x-E, x, x, E-1\}$. The entropy of $\rho(t=\infty,E)$ has a maximum at the point $x=E(1-E)$, which corresponds to a separable state. If the eigenbasis of the operator $A_i$ is rotated relative to the eigenbasis of the excited states occupation operator, the components of the initial state are combinations of components of (\ref{border}). This erases negative dependency, similar to how measuring spin along the rotated axes erases the result of the preceding measurement. Therefore, for $\sigma_x$-interaction, we observe a positive dependency, which nevertheless depends on $E$. Incorporating random phase differences blurs the patterns but does not completely change them.

Finally, let's mention the case when only \textit{one} of the qubits interact. The \textit{dashed} lines in Figure~5d) give dependencies for initial states (\ref{border}) and set the lower boundaries of the allowed regions for initial states satisfying (\ref{eq:ref37c}) for  $\sigma_x$-interaction. For $\sigma_z$ all states satisfying  (\ref{eq:ref37c}), independently of component phases, form segments indicated by dashed lines in Figure~5c) - the final entropy is completely independent of the initial entanglement. This independence also has a place for any size of the system (but only if one qubit interacts). This can be easily shown analytically (see Appendix D).

\begin{figure}[h]
\begin{minipage}[h]{0.44\linewidth}
\center{\includegraphics[width=1\linewidth]{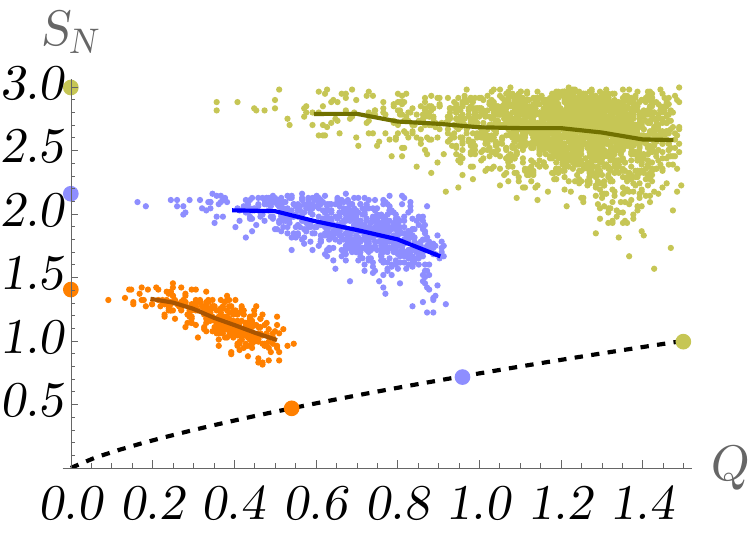}} a) $\sigma_z$-interaction \\
\end{minipage}
\hfill
\begin{minipage}[h]{0.08\linewidth}
\center{\includegraphics[width=1\linewidth]{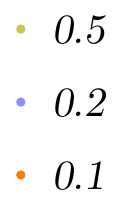}} \\
\end{minipage}
\hfill
\begin{minipage}[h]{0.44\linewidth}
\center{\includegraphics[width=1\linewidth]{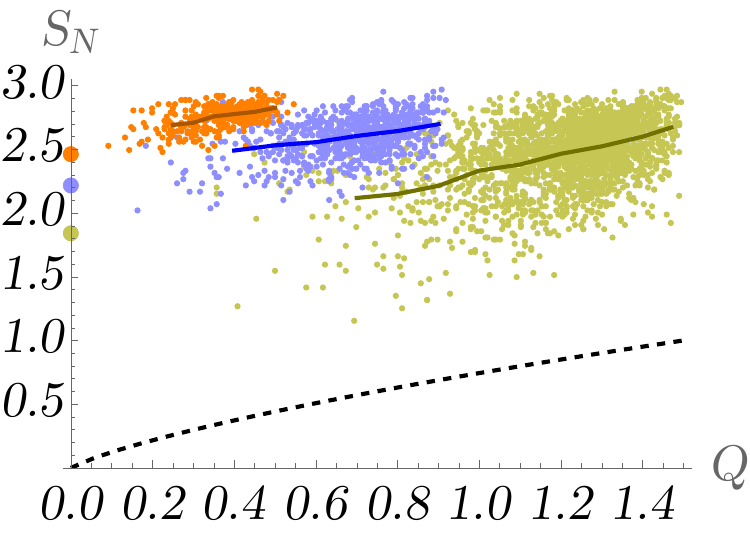}} b) $\sigma_x$-interaction\\
\end{minipage}

\caption{Final entropy vs initial entanglement for random 3-qubit states satisfying condition (\ref{eq:ref37c}). a) Results for $\sigma_z$-interaction, b) results for $\sigma_x$-interaction. Different colors correspond to different values of $E$.  The dashed line shows the boundary of the region allowed for any pure states, in a) it is formed by states $\sqrt{1-E}\ket{000}+\sqrt{E}\ket{111}$.  Solid lines represent segments of approximate average lines for the regions for which a sufficient amount of data has been collected. Bold dots represent a) the points at which the allowed regions touch the dashed line and the $S_N$-axis (the coordinates of these points were obtained analytically), b) Mean points for separately generated separable states.}

\label{Fig.7}
\end{figure}

The results for the 3-qubit system when all qubits interact with the environment are shown in Figure~6. To accumulate sufficient statistics while aiming for the highest possible precision, we rounded $\left\langle {E_i} \right\rangle$ to $0.005, 0.01, 0.02$ for $E=0.5,0.2,0.1$ correspondingly. The chosen approximation slightly expands the boundaries of the regions, especially for small $E$. Nevertheless, the overall dependencies remain the same as for 2 qubits. The main difference is observed in the distribution of points: the points are more strongly concentrated in certain parts of the allowed regions (similarly to Figure~2, this is connected to the concentration of measure).

The considered example illustrates the influence of the sampling on the $S_N(Q)$ dependence. For initial states in which all qubits have the same values of $E$, the patterns noticeably differ from those for Haar-random states. Moreover, the choice of the type of interaction with the environment can change dependence to the opposite. However, the obtained result applies only to small systems. It would be interesting to find specific conditions on the initial states that would significantly alter the dependence for larger systems as well.

\subsection{Generalized Dicke states}
 This subsection is devoted to a non-symmetric generalization of Dicke states, which was used, for example in \cite{Reducible correlations in Dicke states}. We start with a 4-qubit system and the states ($N$ represents the number of excitations):
\begin{equation}
\begin{matrix}
\ket{N=1}=c_1\ket{1000}+c_2\ket{0100}+c_3\ket{0010}+c_4\ket{0001},\\
\ket{N=2}=c_{12}\ket{1100}+c_{13}\ket{1010}+c_{14}\ket{1001}+c_{23}\ket{0110}+c_{24}\ket{0101}+c_{34}\ket{0011}.
\end{matrix}
\label{N1,2}
\end{equation} 
If all coefficients in (\ref{N1,2}) are equal, $\ket{N=1,2}$ are Dicke states and we have $Q(N=1)=1.5, S_z(N=1)=2, Q(N=2)=2, S_z(N=2)=\mathrm{log}_26$. Introducing phase differences between coefficients does not change $S_z$ and $Q$. For $S_z$ it is clear from the expression (\ref{eq:ref39ac}); for $Q$, this can be seen with the help of direct calculation leading to the expressions:

\begin{equation}
\begin{matrix}
\mathrm{for}\ \ket{N=1}: &Q=2(1-\sum_{j}(|c_j|^2)^2),\\
\mathrm{for}\ \ket{N=2}: &Q=2(1-(|c_{12}|^2-|c_{34}|^2)^2-(|c_{13}|^2-|c_{24}|^2)^2-(|c_{14}|^2-|c_{23}|^2)^2).
\end{matrix}
\label{NQ}
\end{equation} 
 Consideration of different absolute values of the coefficients shifts $S_z(N=1,2)$ and $Q(N=1)$ towards smaller values and either lowers or leaves unchanged $Q(N=2)$. Thus, $\sigma_z$-interaction gives positive dependencies of final entropy on initial entanglement, although for $N=2$ it is less pronounced.
 
 Regarding $\sigma_x$-interaction, for Dicke states we have $S_x(N=1)=3,S_x(N=2)=3-\mathrm{log}_43$. If only one coefficient in (\ref{N1,2}) is non-zero, then $S_x(N=1,2)=4$. Incorporating phase differences between coefficients and/or different absolute values shifts $S_x$ toward larger values or leaves it unchanged\footnote{I cannot provide elegant and complete analytical proof here, but this effect is also clearly evident from the numerical investigation.}. Together with a decrease of $Q$, this leads to negative dependencies.

\begin{figure}[h]
\begin{minipage}[h]{0.43\linewidth}
\center{\includegraphics[width=1\linewidth]{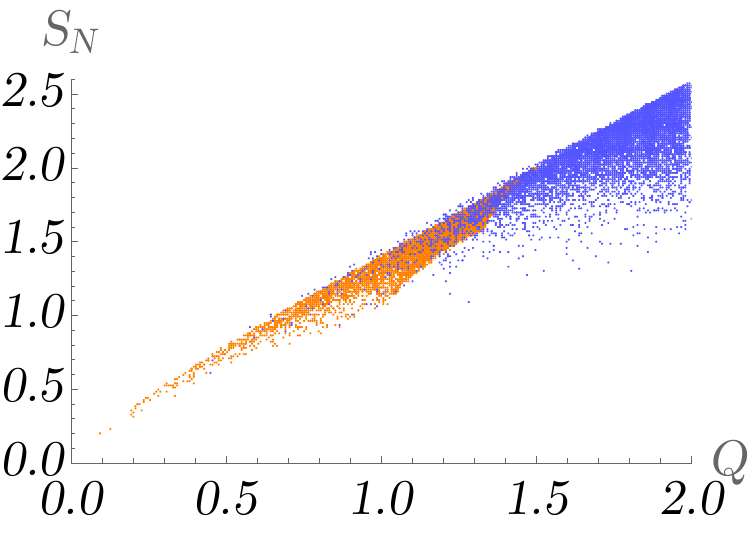}} a) $\sigma_z$-interaction \\
\end{minipage}
\hfill
\begin{minipage}[h]{0.1\linewidth}
\center{\includegraphics[width=1\linewidth]{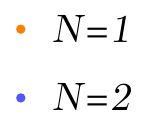}} \\
\end{minipage}
\hfill
\begin{minipage}[h]{0.43\linewidth}
\center{\includegraphics[width=1\linewidth]{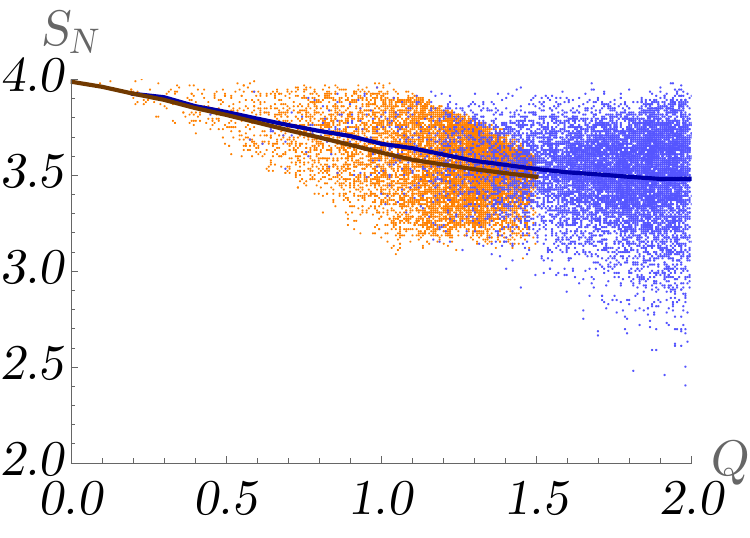}} b) $\sigma_x$-interaction\\
\end{minipage}

\caption{Random states (\ref{N1,2}) exhibit positive dependencies of final entropy on initial entanglement for $\sigma_z$-interaction and negative dependencies for $\sigma_x$. Different colors correspond to different $N$. Approximate average lines are shown for $\sigma_x$-interactions.}

\label{Fig.8}
\end{figure}

Considering $c_j,c_{jk}$ as any complex numbers, satisfying the normalization condition, can be regarded as a non-symmetric generalization of Dicke states. To generate such states randomly, we used expression (\ref{H1}), where for  $N=1$ the vector $\ket{\psi}$ was composed only of $c_j$, and for  $N=2$ only of $c_{jk}$. The resulting Figure~7 confirms our reasoning, showing opposite dependencies for the two types of interactions.

The choice of Dicke states compared to Haar-random ones reduces the effect of measure concentration, which is especially evident for $\sigma_z$-interaction. The larger the $N$ (for the same system size), the less pronounced the dependencies. This can be seen more clearly from the values of the Pearson correlation coefficient and the slope angles of the fitting straight lines, which we provide below for 4-qubit and 6-qubit systems with different numbers of excitations (additionally, we provide values of $\overline{Q}, \overline{S_x}, \overline{S_z}$ and their variations in Appendix E):
 \begin{equation}
			\begin{matrix}
				\ \ \ \ \  \ \ \  & 4q &\ \ & \ \ & \ \ \ \ 6q \\
				\hline
				\ \ \ \ \ \ \ \ &N=1&N=2\ \ \ \ \ \ &N=1&N=2&N=3\\
				\hline
				\mathrm{Angle}(\sigma_x)&-18^\circ&-9^\circ \ \ \ \ &-15^\circ&-4^\circ&-3^\circ\\
				\hline
				\mathrm{Angle}(\sigma_z)&52^\circ&44^\circ \ \ \ \ &58^\circ&49^\circ&47^\circ\\
				\hline
				\mathrm{Correlation} (\sigma_x)&-0.39&-0.16 \ \ \ \ &-0.30&-0.05&0\\
				\hline
				\mathrm{Correlation} (\sigma_z)&0.98&0.82 \ \ \ \ &0.97&0.68&0.60\\
				\end{matrix}	
			\label{data}
	\end{equation} 
 \label{DickePA}

The most remarkable point is that while a small increase in system size noticeably decreases the absolute value of the Pearson coefficient for $\sigma_x$-interaction, the correlation for  $\sigma_z$-interaction decreases much more slowly. For $N=1$, even for relatively large systems, a sufficiently strong correlation\footnote{Although the global dependence is non-linear (which can be seen from Figure~8a)), most values of $S_z$ and $Q$ concentrate within a small region and the Pearson coefficient is close to 1.} will be observed. This is shown in Figure~8. As the system size increases, $\overline{S_z}$ grows and takes the same values as those for Haar-random states with the same number of non-zero components. The average values of $Q$ also increase but tend towards 2. This limitation is explained by the expression (\ref{NQ}) for N=1 which works for any system size.

 \begin{figure}[h]
\begin{minipage}[h]{0.54\linewidth}
\center{\includegraphics[width=1\linewidth]{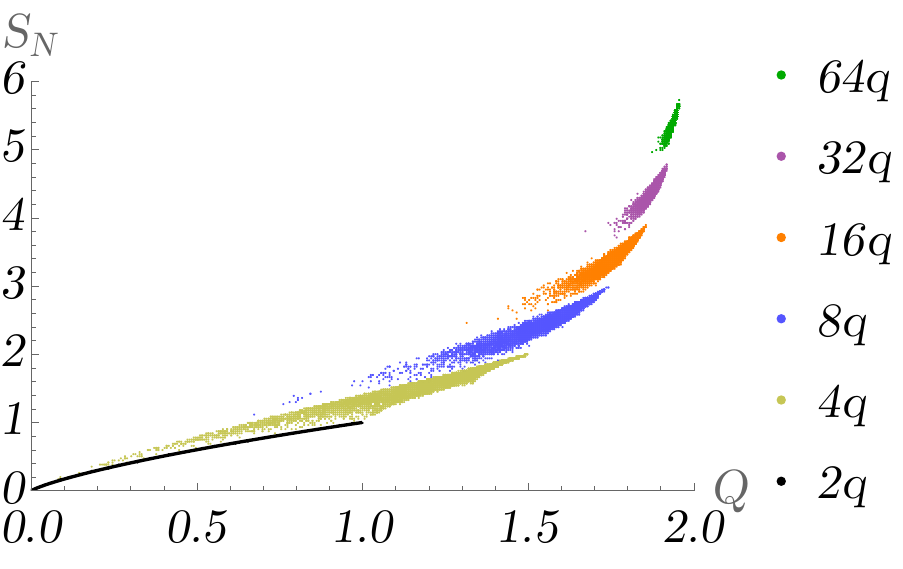}} a) \\
\end{minipage}
\hfill
\begin{minipage}[h]{0.42\linewidth}
	 \begin{equation}
		\begin{matrix}
			\begin{matrix}
				\mathrm{System }\\
				\mathrm{size}
			\end{matrix} & \begin{matrix}
				\mathrm{Correlation}
			\end{matrix} 
			&\begin{matrix}
				\mathrm{Angle}\\
				\end{matrix} \\
			\hline
			2q&1&44^\circ\\
			\hline
			4q&0.98&52^\circ\\
			\hline
			8q&0.96&63^\circ\\
			\hline
			16q&0.95&74^\circ\\
			\hline
			32q&0.94&81^\circ\\
			\hline
			64q&0.93&85^\circ\\
				\end{matrix}
	\end{equation} 
	\center{b)}
\end{minipage}

\caption{The role of the system size for generalized Dicke states with $N=1$ for $\sigma_z$-interaction. a) The bigger the system the larger values takes $S_z$, the closer values of $Q$ are to 2 and the stronger concentrate around their means. b) The increase in system size increases the slope of the fitting straight line, while the Pearson coefficient decreases rather slowly. }

\label{Fig.9}
\end{figure}

This subsection has demonstrated that generalized Dicke states represent another example of sampling, where the choice of interaction type can lead to opposite dependencies of final entropy on initial entanglement. However, negative dependencies were observed only for small systems. At the same time, for $\sigma_z$-interaction, states with $N=1$ show a rather pronounced positive dependency for systems with a sufficiently large number of qubits.

\section{Conclusions} 

In this paper, we investigated the dependence of the entropy increase of a system interacting with the environment on the internal entanglement of the initial state. Even consideration of the simplest model shows that there is no general universal dependency between these two notions. Therefore we examined the issue on average by generating a set of random pure states.

Our study has revealed that regularities are formed by two factors: the shape of the allowed region and the distribution of quantum states within it. For Haar-random states, the allowed regions are rather large, and points within them are distributed in a way that internal entanglement contributes to entropy growth. We observe this dependence for systems of various sizes, both when all qubits interact with the environment and when some qubits are isolated. This is in agreement with the findings of other studies and can be explained by the phenomenon of measure concentration. Additionally, measure concentration is responsible for the fact that a greater depth of entanglement also contributes to the growth of entropy. Simultaneously, imposing additional conditions on the model can significantly constrain the shape of the allowed region and even alter the relationship between final entropy and initial internal entanglement. We observed this for the $\sigma_z$-interaction and initial states with a fixed average excited state occupation of qubits, and also for $\sigma_x$-interaction and generalized Dicke states. The value of average excited state occupation and number of excitations in generalized Dicke states also significantly affect the entropy of the final states.

However, this paper considered only one model. For all interactions in it, the final states of the system did not depend on the environment temperature, which is a drawback from a physics perspective. It would be interesting to explore other types of Hamiltonians (including those for which the environmental state plays a more significant role), compare samples of initial states with various symmetries, study the influence of the initial temperature of the system, pay attention to mixed initial states and initial system-environment correlations, check how the effect of entanglement varies across different stages of evolution.

\section*{Acknowledgements}
I express my gratitude to an anonymous reviewer for reviewing all three versions of my manuscript,  for the efforts and time dedicated to my work, for numerous pieces of advice and comments related to both the physics and the text, as well as for providing references. I also thank Sandu Popescu for a brief conversation that gave rise to the idea for this work, Paweł Horodecki for providing an important reference, and the participants of the Quantum Mathematical Physics seminar at Steklov Mathematical Institute (especially Anton Trushechkin) for valuable discussion.

\appendix
\section*{Appendix A. The proof of ${S_N}(\infty)\geq {S_E}(0)$}
\label{SNE}
\renewcommand{\theequation}{A.\arabic{equation}}
\setcounter{equation}{0}
For pure states, the entanglement entropy of a 2-qubit system is equal to the von Neumann entropy of a single qubit. Therefore to prove ${S_N}(\infty)\geq {S_E}(0)$, we can compare eigenvalues of $\rho_S(\infty)$ (the final density matrix of the entire system) and $\rho_1(0)$ (the initial state of a single qubit).
Any initial pure state of a 2-qubit system can be represented as:
\begin{equation}
		\ket{\psi(0)}=
		\begin{pmatrix}
			\sqrt{\psi_1}e^{i\varphi_1}
			&\sqrt{\psi_2}e^{i\varphi_2}
			&\sqrt{\psi_3}e^{i\varphi_3}
			&\sqrt{\psi_4}e^{i\varphi_4}			
			\end{pmatrix}^T.
			\label{eq:ref65}
	\end{equation} 
We work in the eigenbasis of the interaction operator. If only one qubit interacts, let's say the first one (we can always renumber the qubits), then using (\ref{eq:ref13}) for $\rho_S(\infty)$ we find
\begin{equation}
		\begin{pmatrix}
			\psi_1&\sqrt{\psi_1\psi_2}e^{i(\varphi_1-\varphi_2)}&0&0\\
			\sqrt{\psi_1\psi_2}e^{-i(\varphi_1-\varphi_2)}&\psi_2&0&0\\
			0&0&\psi_3&\sqrt{\psi_3\psi_4}e^{i(\varphi_3-\varphi_4)}\\
			0&0&\sqrt{\psi_3\psi_4}e^{-i(\varphi_3-\varphi_4)}&\psi_4\\		
			\end{pmatrix},
			\label{matr1q2q}
	\end{equation} 
eigenvalues of this matrix are
 \begin{equation}
a_{11}=\psi_1+\psi_2,a_{22}=\psi_3+\psi_4, 0, 0. 
 \label{eq:refeig}
	\end{equation} 
The initial density matrix of the first qubit can be represented as:
\begin{equation}
		M_1(0)=
		\begin{pmatrix}
			a_{11}&a_{12}\\
			a_{21}&a_{22}			
			\end{pmatrix},
			\label{eq:ref66}
	\end{equation} 
 where $a_{11}, a_{22}$ are defined in (\ref{eq:refeig}) and precise form of $a_{12}, a_{21}$ does not matter for our proof. The difference between eigenvalues of $M_1(0)$ is given by 
 \begin{equation}
		\sqrt{(a_{11}-a_{22})^2+4a_{12}a_{21}}.
			\label{eq:ref67}
	\end{equation} 
Since density matrices are Hermitian, $a_{12}a_{21}\geq 0$. Therefore expression (\ref{eq:ref67}) is greater than or equal to the absolute difference of non-zero eigenvalues (\ref{eq:refeig}). A larger difference between eigenvalues implies lower entropy, therefore ${S_N}(\infty)\geq {S_E}(0)$. Thus, we have proven the statement for the case when only one qubit interacts. 

If both qubits interact and condition (\ref{eq:refne}) is fulfilled, the final density matrix has diagonal form with elements $\psi_1,\psi_2,\psi_3,\psi_4$ (to find it we used (\ref{eq:ref38})). Von Neumann entropy expressed through its eigenvalues can be only bigger or equal to the one expressed through eigenvalues (\ref{eq:refeig}). Therefore ${S_N}(\infty)$ in the case when two qubits interact with the environment is greater than or equal to ${S_N}(\infty)$ when only one interacts, and consequently greater than or equal to ${S_E}(0)$.

\section*{Appendix B. Random pure states}
\renewcommand{\theequation}{B.\arabic{equation}}
\setcounter{equation}{0}
\label{random}
For the generation of Haar-random pure states in $l=2^n$-dimensional Hilbert space, we used:
\begin{equation}
\ket{\psi}=
\begin{pmatrix}
			\cos \vartheta_{l-1}\\
			\sin \vartheta_{l-1} \cos \vartheta_{l-2} e^{i \varphi_{l-1}}\\
			\dots\\
			(\prod_{k=2}^{l-1} \sin \vartheta_k) \cos \vartheta_1 e^{i \varphi_2}\\
			(\prod_{k=1}^{l-1} \sin \vartheta_k) e^{i \varphi_1}	
			\end{pmatrix},
			\label{H1}
	\end{equation} 	
where $\vartheta_k=\arcsin(\xi_k^{1/{2k}})$, 
$\xi_k$ are uniformly distributed over the interval $[0,1)$ and  $\varphi_k$ - over $[0, 2\pi)$  \cite{Composed ensembles of random unitary matrices}. Expression (\ref{H1})  can be obtained by applying an arbitrary unitary transformation $U$ (given in Appendix of \cite{Composed ensembles of random unitary matrices}) to the $l$-component vector
 $\begin{pmatrix}
			1&
			0&
			\dots&
			0	
			\end{pmatrix}^T$.

For the generation of random pure separable states $\ket{\Psi}$, we used $n$ random unitary matrices $U_1, \dots, U_n$, uniformly distributed on $SU(2)$:
\begin{equation}
\ket{\psi}=
			U_1\otimes U_2\otimes \dots \otimes U_n 
			\begin{pmatrix}
			1&
			0&
			\dots&
			0	
			\end{pmatrix}^T,
			\label{H2}
	\end{equation} 	
This method and expression (\ref{H1}) for the two-qubit systems are given in Appendix 1 of Ref.~\cite{Dynamics of quantum entanglement}.

\section*{Appendix C. Mean values of $Q$ and $S_N$ for specific values of $E$}
\label{Aver}
The mean values of initial entanglement and final von Neumann entropy $\overline{S}_z$ (for $\sigma_z$-interaction) and $\overline{S}_x$ (for $\sigma_x$-interaction) for 2- and 3-qubit systems with specific values of $E$:
\begin{equation}
	\begin{matrix}
	\begin{matrix}
	\mathrm{2q:}&
			\begin{matrix}
				E&0.5&0.2&0.1&0.05\\
				\hline
				\overline{Q}&0.66&0.35&0.18&0.09\\
				\hline
				\overline{S}_z&1.72&1.27&0.81&0.49\\
				\hline
				\overline{S}_x&1.5&1.66&1.81&1.9.
			\end{matrix}	
\ \ \ \ \ \ \ \ \ \ 
			\end{matrix}&
	\begin{matrix}
		\mathrm{3q:}&
			\begin{matrix}
				E&0.5&0.2&0.1\\
				\hline
				\overline{Q}&1.2&0.68&0.37\\
				\hline
				\overline{S}_z&2.65&1.87&1.16\\
				\hline
				\overline{S}_x&2.46&2.59&2.76.
			\end{matrix}	
			\end{matrix}
	\end{matrix}		
			\label{Mean}
	\end{equation}

 \subsection*{Appendix D. Proof of the independence of the final entropy from the initial entanglement when only one qubit with fixed average occupation number interacts via $\sigma_z$-interaction}
 
 \label{indep}
 \renewcommand{\theequation}{D.\arabic{equation}}
\setcounter{equation}{0}
Any initial pure state of an n-qubit system can be represented in the form ($l=2^n$):
\begin{equation}
		\ket{\psi}=
		\begin{pmatrix}
			\sqrt{\psi_1}e^{i\varphi_1}
			&\dots 
			&  \sqrt{\psi_{l/2}} e^{i\varphi_{l/2}}
			&  \sqrt{\psi_{l/2+1}} e^{i\varphi_{l/2+1}}
			&\dots
			&\sqrt{\psi_l}e^{i\varphi_l}			
			\end{pmatrix}^T.
			\label{eq:ref65b}
	\end{equation} 
 Let only the first qubit interact with the environment. Tracing out all other qubits and using expression (\ref{eq:ref37c}) for the average energy of the first qubit we have:

\begin{equation}
		\left\langle {E_1} \right\rangle=\sum_{k=\frac{l}{2}+1}^{l}\psi_k.	
		\label{eq:ref26}
	\end{equation} 	
On the other hand, entropy is expressed through $\theta_{p,m}$, for which instead of (\ref{eq:ref11b}) we have:
\begin{equation}
	\theta_{p}=|c_{p_1}|^2+|c_{p_2}|^2+\dots+|c_{p_{l/2}}|^2,\ \ \ \theta_{m}=|c_{m_1}|^2+|c_{m_2}|^2+\dots+|c_{m_{l/2}}|^2,	
	\label{eq:ref11d}
	\end{equation} 
where $c_{p_1},\dots,c_{p_{l/2}}, c_{m_1},\dots,c_{m_{l/2}}$ are coefficients in the expansion of the initial state vector in terms of basis vectors
\begin{equation}
 \begin{matrix}
 \ket{p_1}=
	\ket{u_p}
		\otimes
		(\underbrace{
			\begin{matrix}
				1&0&
				\dots&
				0
			\end{matrix}}_{l/2\  components}),\ \ \ \ &\ \ \ 
			\ket{m_1}=
			\ket{u_m}
			\otimes
				(\underbrace{
				\begin{matrix}
					1&0&
					\dots&
					0
				\end{matrix}}_{l/2\ components}),\\
				\ket{p_2}=
	\ket{u_p}
		\otimes
		(\underbrace{
			\begin{matrix}
				0&1&0&
				\dots&
				0
			\end{matrix}}_{l/2\  components}),&\ \ \ \ \ \ \ 
			\ket{m_2}=
			\ket{u_m}
			\otimes
				(\underbrace{
				\begin{matrix}
					0&1&0&
					\dots&
					0
				\end{matrix}}_{l/2\ components}),\\
 \vdots \ \ \ \ & \ \ \vdots\\
	\ket{p_{l/2}}=
			\ket{u_m}
			\otimes
				(\underbrace{
				\begin{matrix}
					0&
					\dots&
					0&
					1
				\end{matrix}}_{l/2\ components}),\ 
				& \ \ \ \ \ket{m_{l/2}}=
		 \ket{u_m}
			\otimes
				(\underbrace{
				\begin{matrix}
					0&
					\dots&
					0&
					1
				\end{matrix}}_{l/2\  components}).
 \end{matrix}
	\label{eq:ref11c}
	\end{equation}

For $\sigma_z$-interaction 
\begin{equation}
\begin{matrix}
		|c_{p_1}|^2=\psi_1, \ \  \dots, \ \   |c_{p_{l/2}}|^2=\psi_{l/2},\\
		 |c_{m_1}|^2=\psi_{l/2+1},\ \ \dots,\ \ |c_{m_{l/2}}|^2=\psi_{l}
		 \end{matrix}
		 \end{equation}
and consequently (\ref{eq:ref11d}) can be expressed as
\begin{equation}
		\theta_p=\sum_{j=1}^{\frac{l}{2}}\psi_j	=1-\left\langle {E_1} \right\rangle, \ \ \ 
		\theta_m=\sum_{j=\frac{l}{2}+1}^{l}\psi_j=\left\langle {E_1} \right\rangle.
		\label{eq:ref27}
	\end{equation} 	
Therefore entropy for fixed $\left\langle {E_1} \right\rangle$ is constant.

 \subsection*{Appendix E. Mean values for generalized Dicke states}
 Below we provide means and variations for generalized Dicke states with different number of excitations N for 4 and 6-qubit systems. For comparison, we include in the table corresponding quantities for Haar-random (H-R) states.
 \begin{equation}
			\begin{matrix}
				\ \ \ \ \  \ \ \ &\ \ & 4q &\ \ & \ \ & \ \ \ \ 6q \\
				\hline
				\ \ \ \ \ \ \ \ &H-R&N=1&N=2\ \ \ \ \ \ &H-R&N=1&N=2&N=3\\
				\hline
				\begin{matrix}
				\mathrm{Number\ of\ free}\\
				\mathrm{components}
			\end{matrix}&16&4&6  \ \ \ \ &64&6 &15&20\\
				\hline
				\overline{Q}&1.65&1.20&1.71 \ \ \ \ &2.86&1.43&2.50&2.86\\
				\hline
				\overline{S_x}&3.43&3.56&3.51 \ \ \ \ &5.40&5.51&5.44&5.43\\
				\hline
				\overline{S_z}&3.43&1.56&2.09 \ \ \ \ &5.40&2.09&3.35&3.75\\
				\hline
				Var(Q)&0.024&0.046&0.045 \ \ \ \ &0.002&0.022&0.011&0.009\\
				\hline
				Var(S_x)&0.031&0.029&0.045 \ \ \ \ &0.009&0.018&0.022&0.024\\
				\hline
				Var(S_z)&0.031&0.076&0.064 \ \ \ \ &0.009&0.064&0.033&0.026\\
			\end{matrix}	
			\label{data}
	\end{equation} 
 \label{Dicke}

\end{document}